\documentclass[doublespace,11pt]{article}
\usepackage{graphicx}
\usepackage{setspace}
\usepackage{mathptm}
\usepackage{amsmath, amsthm}
\usepackage{subfigure}
\usepackage{epsfig}
\usepackage{url}

\begin{document}
\sloppy
\begin{spacing}{1}

\begin{titlepage}
\hspace{0.08in}
\begin{minipage}{\textwidth}
\begin{center}
\vspace*{3cm}
\begin{tabular}{c c c}
\hline
 & & \\
 & {\Huge The University of Algarve} & \\
 & & \\
 & {\Huge Informatics Laboratory} & \\
 & & \\
\hline
\end{tabular}\\
\vspace*{2cm}
{\Large
UALG-ILAB\\
Technical Report No. 200705 \\
June, 2007\\
}
\vspace*{3cm}
{\bf EasyVoice: Integrating voice synthesis with Skype}\\
\addvspace{0.5in}
{\bf Paulo A. Condado} and {\bf Fernando G. Lobo}\\
\vspace*{-0.1in}
\vspace*{4cm}
Department of Electronics and Informatics Engineering\\
Faculty of Science and Technology \\
University of Algarve \\
Campus de Gambelas\\
8000-117 Faro, Portugal\\
URL: {\verb http://www.ilab.ualg.pt }\\
Phone: (+351) 289-800900\\
Fax: (+351) +351 289 800 002 \\
\end{center}
\end{minipage}
\end{titlepage}

\title{\bf EasyVoice: Integrating voice synthesis with Skype}
\author{    
            {\bf Paulo A. Condado}\\
            \small UAlg Informatics Lab\\
            \small DEEI-FCT, University of Algarve\\
            \small Campus de Gambelas\\
            \small 8000-117 Faro, Portugal\\
            \small pcondado@ualg.pt
\and
          {\bf Fernando G. Lobo}\\
            \small UAlg Informatics Lab\\
            \small DEEI-FCT, University of Algarve\\
            \small Campus de Gambelas\\
            \small 8000-117 Faro, Portugal\\
            \small flobo@ualg.pt
}
\date{}
\maketitle

\begin{abstract}

This paper presents {\em EasyVoice}, a system that integrates voice
synthesis with Skype. EasyVoice allows a person with voice
disabilities to talk with another person located anywhere in the
world, removing an important obstacle that affect these people
during a phone or VoIP-based conversation.
\end{abstract}

\section{Introduction}
\label{sec:intro}


Voice over IP (VoIP) applications have become very popular in recent
years. These applications allow people to talk for free over the
Internet and also to make traditional calls through the
Public-Switched Telephone Network at a small fraction of the cost
offered by traditional phone companies. One of the most popular VoIP
applications is Skype, which is freely available at
\url{http://www.skype.com}.

People with voice disabilities, however, are oftentimes not able to
use a VoIP application (or, as a matter of fact, a regular or mobile
phone) in order to have a conversation with another person. A minor
voice disability might not be a big obstacle, but for those with
severe voice disabilities, a phone conversation is something almost
impossible to achieve. For those cases, the only hope appears to be
the utilization of a text-to-speech (TTS) system together with a
VoIP application. This solution has been hinted by ourselves in the
recent past~\cite{Condado:06} and is something that to our best
knowledge has not been tried before.

The utilization of speech synthesis~\cite{Dutoit:2003} as an
assistive technology allows barriers to be removed for people with a
wide range of disabilities. Application examples include the use of
screen-readers for people with visual impairment, as well as to help
people with dyslexia and other reading difficulties. Another
important utilization, as we are about to see in this paper, is to
aid those with severe voice disabilities.


\section{The EasyVoice System}
\label{sec:easyvoice-tts}

\subsection{General idea}

With text-to-speech, a person types at a keyboard, the text is
synthesized, and the sound comes out through the computer speakers.
With EasyVoice, the sound is injected directly through the network
rather than being sent to the computer speakers. EasyVoice achieves
this by working together with Skype via its Application Programming
Interface (API) (see \url{https://developer.skype.com/}).

The current implementation of EasyVoice works under Microsoft
Windows. It is freely available at
\url{http://w3.ualg.pt/~pcondado/easyvoice/}. Any speech synthesizer
can be used as long as it is SAPI 5 compliant. SAPI stands for
Speech Application Programming Interface. It is an API developed by
Microsoft to allow the use of Speech Recognition and Speech
Synthesis within Windows applications.

Notice that one could think that there is no need to do any special
integration between a speech synthesizer and a VoIP application.
After all, one could simply use a speech synthesizer and let the
computer's microphone be close enough to the computer speakers. Such
a naive solution, however, yields an excessive amount of echo during
the conversation because the person at the other end of the line
hears back her own voice.

\subsection{Features for slow typers}

Many times, a person with voice disabilities has also motor
coordination problems, and it may happen that the person types very
slowly. That's often the case of people with cerebral palsy. In such
cases, typing with a reasonable speed at a regular keyboard can be a
difficult thing to do, and that is an obstacle for a smooth phone
conversation.

To alleviate this problem, EasyVoice provides a number of features
which can be used to accelerate the typing process. Although they
can be very useful for slow typers, they may not be useful at all
for those with a normal typing ability. Therefore, all the features
are optional and can be turned on/off by pressing a single button.
The features implemented are the following:

\begin{itemize}
\item archive of recent messages
\item word completion
\item abbreviation system
\item virtual keyboard
\end{itemize}

\begin{figure}
\center
\epsfig{file=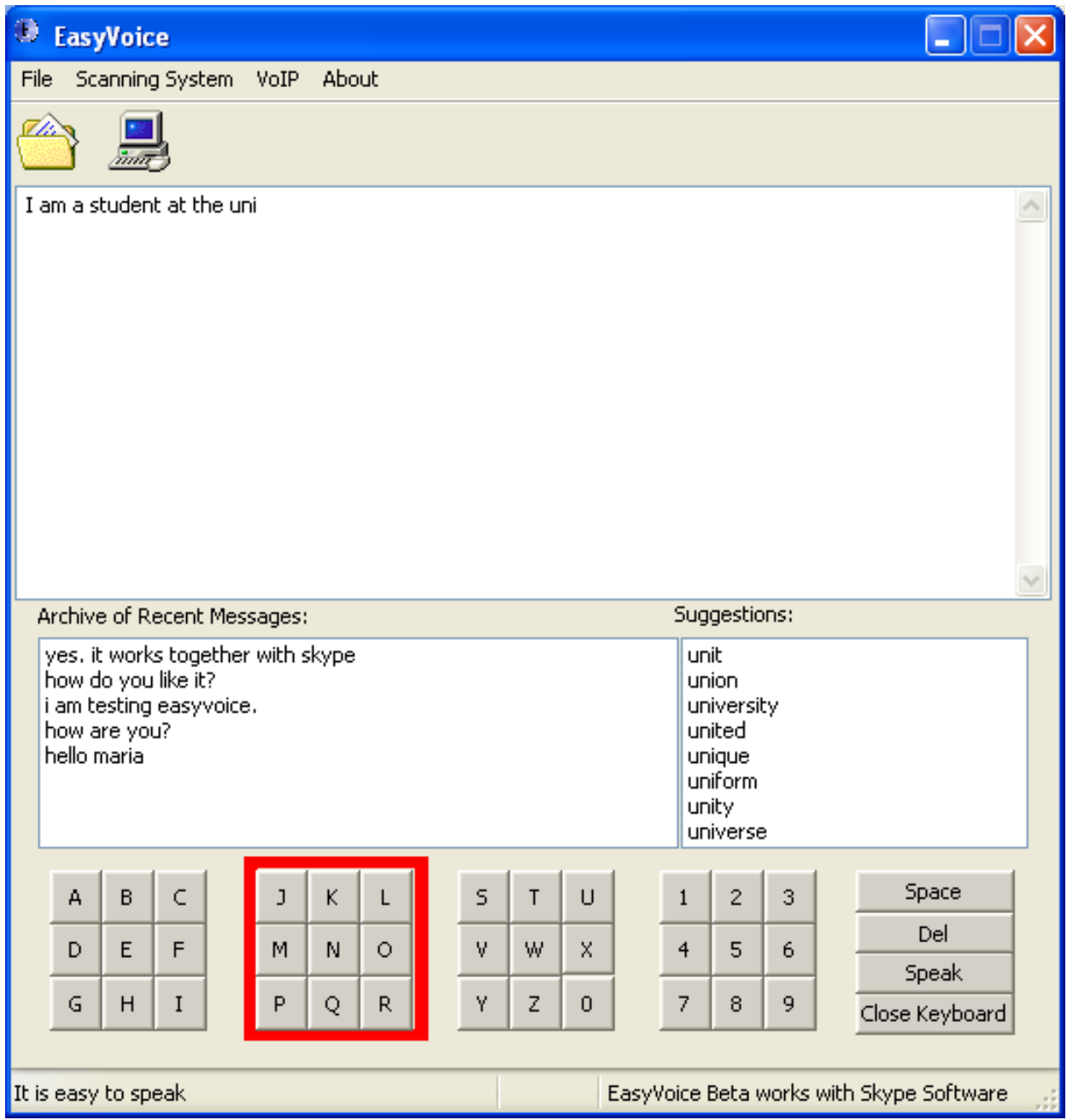,width=1.0\columnwidth}
\caption{The EasyVoice user interface. The top part shows a text
panel for input. In the middle part, the left panel has the archive
of recent messages, and the right panel shows the suggestions given
by the word completion algorithm. The bottom part shows the virtual
keyboard with its scanning system in operation. }
\label{fig:easyvoice-with-keyboard}
\end{figure}

Figure~\ref{fig:easyvoice-with-keyboard} shows a screenshot of the
EasyVoice user interface with all the features turned on. There is a
text panel on the top part where the user can type. Below the text
panel there are two list boxes. The one on the left part is a list
of recently typed messages, and on the right part is a list of
possible words given by the word completion algorithm.

The archive of recently typed messages is useful because during a
conversation it is many times necessary to repeat some words or
phrases. With the archive in hand, the user does not need to retype
the message and can simply pick it from the list again.

The word completion algorithm searches in a built-in dictionary for
those words that have as a prefix the sequence of letters typed by
the user so far. The system gives a list of the 8 most frequent
words of the language, which were obtained by the British National
Corpus (see \url{http://www.kilgarriff.co.uk/bnc-readme.html}).

Another important feature is the abbreviation system. It is common
for people to use abbreviations when writing. For example, in
English it is common for people to use ``{\tt btw}'' as an
abbreviation of ``{\tt by the way}''. Within EasyVoice, the user can
define his own abbreviations and the system automatically replaces
each abbreviation by the corresponding words, before sending it to
the speech synthesizer.

The final feature is a virtual keyboard with a scanning system
incorporated. Many individuals cannot control their hands with
enough accuracy to use a regular computer keyboard, and sometimes
only have the ability to control a single touch button. Virtual
keyboards are a reasonable solution to solve some of these
limitations. With a scanning system, a set of options is presented
to the user on the computer screen, and a visual cursor advances
through the options, one at time, at a specified time rate. The user
responds by pressing a touch button whenever the cursor is on top of
the desired option. Sometimes an option is just a container for more
options and is referred to as a {\em group option}. When a group
option is selected, the scanning system immediately focuses on the
sub-options of that group, and again, advances the visual cursor
through each of them~\cite{Demasco:1992}.

\section{Conclusions}
\label{sec:conclusions} This paper presented EasyVoice, a system
that combines existing technologies in a novel way for helping
people with voice disabilities.

An innovation oftentimes emerge from the combination of ideas. A
good example is the creation of the World Wide Web by Tim
Berners-Lee. The Web came upon existence by joining two technologies
that already existed for quite some time: (1) the TCP/IP protocol
suite, and (2) the notion of Hypertext. What Berners-Lee did,
according to his own words, was to marry the two notions
together~\cite{Berners-Lee:1999}.

EasyVoice has in itself many things in common with the creation of
the Web. User interfaces for people with disabilities, speech
synthesizers, and VoIP applications, are all technologies that
already existed for quite some time. What we did was to marry them
together, and by doing that, we believe we have created an
innovation, something that did not exist before and that opens a
window for a new world of communications, learning, and
socialization, for people with voice disabilities.

\section{Acknowledgments}
This work was sponsored by Funda\c{c}\~{a}o Caloust Gulbenkian under
grant Proc.~65538, and also by the Portuguese Foundation for Science
and Technology (FCT/MCTES) under grant \text{POCI/CED/62497/2004}.

%

%
%

\end{spacing}
\end{document}